\documentclass[12pt]{iopart}

\usepackage{iopams}

\def  \be   {\begin{equation}}
\def  \ee   {\end{equation}}
\def  \beq  {\begin{eqnarray}}
\def  \eeq  {\end{eqnarray}}

\begin{document}

\title[Fast and slow two-fluid magnetic reconnection]
{Fast and slow two-fluid magnetic reconnection}

\author{Leonid M. Malyshkin}

\address{Department of Astronomy and Astrophysics,
University of Chicago, 5640 S. Ellis Ave., Chicago, IL 60637}

\ead{leonmal@uchicago.edu}

\begin{abstract}
We present a two-fluid magnetohydrodynamics (MHD) model of 
quasi-stationary, two-dimensional magnetic reconnection in 
an incompressible plasma composed of electrons and ions. 
We find two distinct regimes of slow and fast reconnection. 
The presence of these two regimes can provide a possible 
explanation for the initial slow build up and subsequent 
rapid release of magnetic energy frequently observed in
cosmic and laboratory plasmas.
\end{abstract}  

%52.35.Vd 	Magnetic reconnection in plasmas
%94.30.cp 	Magnetic reconnection (Physics of the magnetosphere)
%96.60.Iv 	Magnetic reconnection (Solar physics)
%95.30.Qd       Plasmas astrophysical
%52.27.Cm       Plasmas multicomponent

\maketitle

%---------------------------------------------------------------------

\section{\label{INTRODUCTION}
Introduction
}
Magnetic reconnection is the physical process by means of which magnetic 
field lines join one another and rearrange their topology. Magnetic 
reconnection is believed to be the mechanism by which magnetic energy 
is converted into kinetic and thermal energy in the solar atmosphere, 
the Earth's magnetosphere, and in laboratory 
plasmas~\cite{biskamp_2000,priest_2000,bc_low_2003,kulsrud_2005,drake_2006,
yamada_2009,zweibel_2009}.
Many reconnection related physical phenomena observed in cosmic and
laboratory plasmas exhibit a two-stage behavior. During the first stage,
magnetic energy is slowly built up and stored in the system with
relatively little reconnection occurring. The second stage is characterized
by a sudden and rapid release of the accumulated magnetic energy due to
a fast reconnection process. For example, a solar flare is powered by a
sudden (on timescale ranging from minutes to tens of minutes) release of 
magnetic energy stored in the upper solar atmosphere~\cite{kulsrud_2005}.
Because the value of the Spitzer electrical resistivity is very low in 
hot plasmas, magnetic energy release rates predicted by a simple 
single-fluid MHD description of magnetic reconnection are much slower 
than the rates observed during fast reconnection events in astrophysical 
and laboratory plasmas~\cite{biskamp_2000,bc_low_2003,kulsrud_2005,drake_2006,
yamada_2009,zweibel_2009}. 
One of the most promising solutions of this discrepancy is the 
two-fluid MHD theoretical approach to magnetic 
reconnection~\cite[and references therein]{biskamp_2000,
kulsrud_2005,drake_2006,yamada_2009,zweibel_2009}.
Recently a model of two fluid reconnection in a electron-proton plasma 
was presented in~\cite{malyshkin_2009}. In this paper, we consider a 
more general case of two-fluid reconnection in electron-ion and 
electron-positron plasmas, and we present derivations in detail.
In the discussion section, we also argue that 
the slow and fast reconnection regimes predicted by our model, can provide 
a possible explanation for the observed two-stage reconnection behavior.

%---------------------------------------------------------------------

\section{\label{MHD_EQUATIONS}
Two-fluid MHD equations
}
In this study, we use physical units in which the speed of light $c$ 
and four times $\pi$ are replaced by unity, $c=1$ and $4\pi=1$. To 
rewrite our equations in the Gaussian centimeter-gram-second (CGS) 
units, one needs to make the following substitutions: magnetic field 
${\bf B}\rightarrow {\bf B}/\sqrt{4\pi}$, electric field 
${\bf E}\rightarrow c{\bf E}/\sqrt{4\pi}$, electric current 
${\bf j}\rightarrow\sqrt{4\pi}\,{\bf j}/c$, electrical resistivity 
$\eta\rightarrow\eta c^2/4\pi$, the proton electric charge 
$e\rightarrow\sqrt{4\pi}\,e/c$.

We consider an incompressible two-component plasma, composed of 
electrons and ions. We assume the plasma is non-relativistic and, 
therefore, quasi-neutral. The ions are assumed to have
mass $m_i$ and electric charge $Ze$, while the electrons have 
mass $m_e$ and charge $-e$. Because of incompressibility, the 
electron and ion number densities are constant, 
\beq
n_e\equiv n={\rm const}, \qquad n_i=Z^{-1}n={\rm const},
\label{NUMBER_DENSITIES}
\eeq
where the last formula follows from the plasma quasi-neutrality condition
$Zen_i=en_e$. The plasma density $\rho$, the electric current ${\bf j}$
and the plasma (center-of-mass) velocity ${\bf V}$ are 
\beq
\rho &=& m_in_i+m_en_e=n(Z^{-1}m_i+m_e)={\rm const},
\label{DENSITY_INITIAL}
\\
{\bf j} &=& Zen_i{\bf u}^i-en_e{\bf u}^e=ne({\bf u}^i-{\bf u}^e),
\label{CURRENT}
\\
{\bf V} &=& (m_in_i{\bf u}^i+m_en_e{\bf u}^e)/\rho=n(Z^{-1}m_i{\bf u}^i+m_e{\bf u}^e)/\rho.
\label{PLASMA_V}
\eeq 
Here ${\bf u}^e$ and ${\bf u}^i$ are the mean electron and ion velocities,
which can be found from the above equations,
\beq
{\bf u}^e={\bf V}-(m_i/Ze\rho){\bf j}, \qquad
{\bf u}^i={\bf V}+(m_e/e\rho){\bf j}.
\label{E_I_VELOCITIES}
\eeq

The equations of motion for the electrons and ions 
are~\cite{braginskii_1965,sturrock_1994} 
\begin{eqnarray}
n_em_e\left[\partial_t{\bf u}^e
+({\bf u}^e{\bf\nabla}){\bf u}^e\right]&=&-{\bf\nabla}P_e
-n_ee({\bf E}+{\bf u}^e\times{\bf B})-{\bf K},
\label{MOTION_E_INITIAL}
\\
n_im_i\left[\partial_t{\bf u}^i
+({\bf u}^i{\bf\nabla}){\bf u}^i\right]&=&-{\bf\nabla}P_i
+n_iZe({\bf E}+{\bf u}^i\times{\bf B})+{\bf K},
\label{MOTION_P_INITIAL}
\end{eqnarray}
where $P_e$ and $P_i$ are the electron and ion pressure tensors, and 
${\bf K}$ is the resistive frictional force due to electron-ion 
collisions. Force ${\bf K}$ can be approximated 
as~\cite{braginskii_1965,sturrock_1994} 
\beq
{\bf K}=n^2e^2\eta({\bf u}^e-{\bf u}^i)=-ne\eta{\bf j}, 
\label{K_FORCE}
\eeq
where $\eta$ is the electrical resistivity, and we use equation~(\ref{CURRENT}). 
For simplicity, we assume isotropic resistivity, and we also neglect ion-ion 
and electron-electron collisions and the corresponding viscous forces. 
Substituting equations~(\ref{NUMBER_DENSITIES}),~(\ref{E_I_VELOCITIES})
and~(\ref{K_FORCE}) into equations~(\ref{MOTION_E_INITIAL}) 
and~(\ref{MOTION_P_INITIAL}), we obtain
\begin{eqnarray}
nm_e\left[\partial_t{\bf V}+({\bf V}{\bf\nabla}){\bf V}\right]
\nonumber\\
\qquad -(nm_em_i/Ze\rho)\left[\partial_t{\bf j}+({\bf V}{\bf\nabla}){\bf j}
+({\bf j}{\bf\nabla}){\bf V}-(m_i/Ze\rho)({\bf j}{\bf\nabla}){\bf j}\right]
\nonumber\\
\;\; {}=-{\bf\nabla}P_e-ne{\bf E}-ne{\bf V}\times{\bf B}+(m_in/Z\rho){\bf j}\times{\bf B}
+ne\eta{\bf j},
\label{MOTION_E}
\\
Z^{-1}nm_i\left[\partial_t{\bf V}+({\bf V}{\bf\nabla}){\bf V}\right]
\nonumber\\
\qquad +(nm_em_i/Ze\rho)\left[\partial_t{\bf j}+({\bf V}{\bf\nabla}){\bf j}
+({\bf j}{\bf\nabla}){\bf V}+(m_e/e\rho)({\bf j}{\bf\nabla}){\bf j}\right]
\nonumber\\
\;\; {}=-{\bf\nabla}P_i+ne{\bf E}+ne{\bf V}\times{\bf B}+(m_en/\rho){\bf j}\times{\bf B}
-ne\eta{\bf j}.
\label{MOTION_P}
\end{eqnarray}
We sum equations~(\ref{MOTION_E}) and~(\ref{MOTION_P}) together and obtain the 
plasma momentum equation 
\beq
\rho\left[\partial_t{\bf V}+({\bf V}{\bf\nabla}){\bf V}\right]
+(m_em_i/Ze^2\rho)({\bf j}{\bf\nabla}){\bf j}=
-{\bf\nabla}P+{\bf j}\times{\bf B},
\label{MOTION_LAW_INITIAL}
\eeq
where $P=P_e+P_i$ is the total pressure. Next we subtract 
equation~(\ref{MOTION_P}) multiplied by $Zm_e/m_i$ 
from equation~(\ref{MOTION_E}) and obtain the generalized Ohm's law
\beq
{\bf E} &=&
\eta{\bf j}-{\bf V}\times{\bf B}+(m_i/Ze\rho)(1-Zm_e/m_i){\bf j}\times{\bf B}
\nonumber
\\
&& -(m_i/Ze\rho)[{\bf\nabla}P_e-(Zm_e/m_i){\bf\nabla}P_i]
\nonumber
\\
&& +(m_em_i/Ze^2\rho)\left[\partial_t{\bf j}+({\bf V}{\bf\nabla}){\bf j}
+({\bf j}{\bf\nabla}){\bf V}\right.
\nonumber
\\
&& \qquad\qquad \left.-(m_i/Ze\rho)(1-Zm_e/m_i)({\bf j}{\bf\nabla}){\bf j}\right].
\label{OHMS_LAW_INITIAL}
\end{eqnarray}
It is convenient to introduce the ion and electron inertial lengths
\beq
\begin{array}{l}
d_i\equiv(m_i/n_iZ^2e^2)^{1/2}=(m_i/Zne^2)^{1/2},\\
d_e\equiv(m_e/n_ee^2)^{1/2}=(m_e/ne^2)^{1/2}\le d_i, 
\end{array}
\label{INERTIAL_LENGTHS}
\eeq
and constants
\beq
\begin{array}{l}
\omega_+^2\equiv(1+Zm_e/m_i)^{-1}=(1+d_e^2/d_i^2)^{-1}, \\
\omega_-^2\equiv 1-Zm_e/m_i=1-d_e^2/d_i^2\ge 0.
\end{array}
\label{OMEGAS}
\eeq
Here we consider a physically relevant case of $Zm_e\le m_i$, so 
that $d_e\le d_i$, $0\le\omega_-^2<1$ and $1/2\le\omega_+^2<1$. 
Note that $\omega_+^2\approx \omega_-^2\approx 1$ in the case of 
electron-ion plasma ($Zm_e\ll m_i$), and $\omega_+^2=1/2$ and $\omega_-^2=0$ 
in the case of electron-positron plasma ($Z=1$ and $m_i=m_e$). 

Using definitions~(\ref{INERTIAL_LENGTHS}) and~(\ref{OMEGAS}), we
obtain for the plasma density~(\ref{DENSITY_INITIAL}) expression
\beq
\rho=m_in/Z\omega_+^2=n^2e^2d_i^2/\omega_+^2, 
\label{DENSITY}
\eeq
and we rewrite the plasma momentum equation~(\ref{MOTION_LAW_INITIAL}) 
and Ohm's law~(\ref{OHMS_LAW_INITIAL}) as
\beq
&&\rho\left[\partial_t{\bf V}+({\bf V}{\bf\nabla}){\bf V}\right]
+\omega_+^2d_e^2({\bf j}{\bf\nabla}){\bf j}=
-{\bf\nabla}P+{\bf j}\times{\bf B},
\label{MOTION_LAW}
\\
&&{\bf E}=\eta{\bf j}-{\bf V}\!\times\!{\bf B}
+(\omega_+^2\omega_-^2/ne)\,{\bf j}\!\times\!{\bf B}
\nonumber
\\
&& \hphantom{{\bf E}=} 
-(\omega_+^2/ne)[{\bf\nabla}P_e-(d_e^2/d_i^2){\bf\nabla}P_i]
\nonumber
\\
&& \hphantom{{\bf E}=} 
+\omega_+^2d_e^2\left[\partial_t{\bf j}+({\bf V}{\bf\nabla}){\bf j}
+({\bf j}{\bf\nabla}){\bf V}
-(\omega_+^2\omega_-^2/ne)({\bf j}{\bf\nabla}){\bf j}\right]\!.
\label{OHMS_LAW}
\end{eqnarray}
It is noteworthy that the electron 
inertia terms, proportional to $d_e^2$, enter both Ohm's 
law and the momentum equation. Although these terms are important
for fast two-fluid reconnection (as we shall see below), they
have been frequently neglected in the momentum equation in the 
past~\footnote{
For particle species $s\in\{e,i\}$ we use the standard definition 
of the pressure tensor as the density times the second moment of 
the particles velocity fluctuations relative to the mean velocity, 
$P_s\equiv n_sm_s\langle({\bf \upsilon}^s
-{\bf u}^s)({\bf \upsilon}^s-{\bf u}^s)\rangle$, 
where ${\bf u}^s=\langle{\bf \upsilon}^s\rangle$~\cite{braginskii_1965}. 
Instead, one could use velocity fluctuations relative to the plasma 
center-of-mass velocity~(\ref{PLASMA_V}) and define pressure as
${\tilde P}_s\equiv n_sm_s\langle({\bf \upsilon}^s
-{\bf V})({\bf \upsilon}^s-{\bf V})\rangle~$~\cite{sturrock_1994}.
In this case, the total pressure tensor would be
${\tilde P}={\tilde P}_e+{\tilde P}_i=P+\omega_+^2d_e^2\,{\bf j}\,{\bf j}$, 
and, therefore, the electron inertia term 
$\omega_+^2d_e^2({\bf j}{\bf\nabla}){\bf j}$ in the momentum 
equation~(\ref{MOTION_LAW}) would become absorbed into the pressure 
term ${\bf\nabla}{\tilde P}$. However, note that pressure ${\tilde P}$ 
is strongly anisotropic.
}. 
In addition, we note that ${\bf\nabla}\cdot{\bf B}=0$, and also 
${\bf\nabla}\cdot{\bf V}=0$ and ${\bf\nabla}\cdot{\bf j}=0$ for 
incompressible and non-relativistic plasmas.

For convenience of the presentation, below we will refer to the plasma as being 
electron-ion, even though, unless otherwise stated, our derivations in the 
next two sections are valid for reconnection in an electron-positron plasma 
as well.

%---------------------------------------------------------------------

\section{\label{LAYER}
Reconnection layer
}
\begin{figure}[t]
\vspace{3.3truecm}
\includegraphics{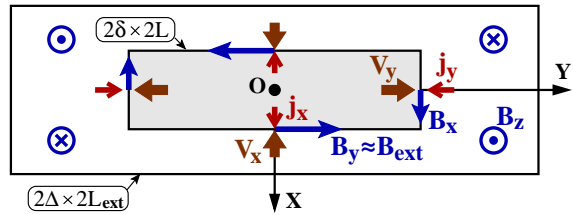}
\caption{Two-dimensional reconnection layer.
}
\label{FIGURE_LAYER}
\end{figure}

We consider two-fluid magnetic reconnection in the classical 
two-dimensional Sweet-Parker-Petschek geometry, which is shown in
figure~\ref{FIGURE_LAYER}. The reconnection layer is in the $x$-$y$ 
plane with the $x$- and $y$-axes perpendicular to and along the 
reconnection layer respectively. The $z$ derivatives of all physical 
quantities are zero. 

The approximate thickness of the reconnection current layer is $2\delta$, 
which is defined in terms of the out-of-plane current ($j_z$) profile 
across the layer~\footnote{Thickness $\delta$ can be formally defined 
by fitting the Harris sheet profile $(B_{ext}/\delta)cosh^{-2}(x/\delta)$ 
to the current profile $j_z(x,y=0)$.
}. 
The approximate length of the out-of-plane current ($j_z$) profile along 
the layer is defined as $2L$. 
Outside the reconnection current layer the electric currents are weak, the 
electron inertia is negligible, Ohm's law~(\ref{OHMS_LAW}) reduces 
to ${\bf E}=-{\bf V}\times{\bf B}+{\bf j}\times{\bf B}/ne
=-{\bf u^e}\times{\bf B}$ (in the case of electron-ion plasma, 
$\omega_+^2\approx \omega_-^2\approx 1$), and, therefore, the magnetic 
field lines are frozen into the electron fluid. Thus, $2\delta$ and 
$2L$ are also approximately the thickness and the length of the 
electron layer, where electron inertia is important and the electrons 
are decoupled from the field lines. 
The ion layer, where the ions are decoupled from the field lines, 
is assumed to have thickness $2\Delta$ and length $2L_{ext}$, which can 
be much larger than $2\delta$ and $2L$ respectively. 
The values of the reconnecting field in the upstream regions outside 
the electron layer (at $x\approx\delta$) and outside the ion layer 
(at $x\approx\Delta$) are about the same, $B_y\approx B_{ext}$ up to a 
factor of order unity. This result follows directly from the definition 
of $2\delta$, and from the $z$-component of the Ampere's law, 
$B_y(x,y=0)=\int_0^x j_z(x',y=0)dx'$.
The out-of-plane field $B_z$ is assumed to have a quadrupole structure (see 
figure~\ref{FIGURE_LAYER})~\cite{drake_2006,yamada_2009,zweibel_2009}~\footnote{
Below we shall see that $B_z$ has quadrupole structure only in the case of 
electron-ion plasma, but not in the case of electron-positron plasma.
}. 

The reconnection layer is assumed to have a point symmetry with 
respect to its geometric center~$O$ (see figure~\ref{FIGURE_LAYER}) 
and reflection symmetries with respect to the $x$- and $y$-axes. 
Thus, the $x$-, $y$- and $z$-components of ${\bf V}$, ${\bf B}$ and 
${\bf j}$ have the following symmetries:
$V_x(\pm x,\mp y)=\pm V_x(x,y)$, $V_y(\pm x,\mp y)=\mp V_y(x,y)$,
$V_z(\pm x,\mp y)=V_z(x,y)$,
$B_x(\pm x,\mp y)=\mp B_x(x,y)$, $B_y(\pm x,\mp y)=\pm B_y(x,y)$,
$B_z(\pm x,\mp y)=-B_z(x,y)$,
$j_x(\pm x,\mp y)=\pm j_x(x,y)$, $j_y(\pm x,\mp y)=\mp j_y(x,y)$
and $j_z(\pm x,\mp y)=j_z(x,y)$. The derivations below extensively 
exploit these symmetries and are similar to the derivations 
in~\cite{malyshkin_2009,malyshkin_2005,malyshkin_2008}.

We make the following assumptions for the reconnection process. 
First, resistivity $\eta$ is assumed to be constant and very small, 
so that the characteristic Lundquist number $S$ is very large,
\beq
S\equiv V_AL_{ext}/\eta \gg 1, \qquad V_A\equiv B_{ext}/\sqrt{\rho}.
\label{S_and_V_A}
\eeq
Here $V_A$ is the Alfven velocity. 
Second, the reconnection process is assumed to be quasi-stationary
(or stationary), so that we can neglect time derivatives in 
the equations above and in the derivations below. This assumption is 
satisfied if there are no plasma instabilities in the reconnection layer, 
and the reconnection rate is slow sub-Alfvenic, $E_z\ll V_AB_{ext}$. 
Third, we assume that the reconnection layer is 
thin, $\delta\ll L$ and $\Delta\ll L_{ext}$, which is an assumption 
related to the previous one.
Fourth, we assume that the electron and ion pressure tensors 
$P_e$ and $P_i$ are isotropic, therefore, the pressure terms in 
equations~(\ref{OHMS_LAW}) and~(\ref{MOTION_LAW}) are assumed to 
be scalars. 

%---------------------------------------------------------------------

\section{\label{RECONNECTION_EQUATIONS}
Two-fluid reconnection equations
}
We use Ampere's law and neglect the displacement current in a 
non-relativistic plasma to find the components of the electric current 
\beq
j_x=\partial_y B_z, \quad 
j_y=-\partial_x B_z, \quad 
j_z=\partial_x B_y-\partial_y B_x.
\label{J_DERIVATIVES}
\eeq 
The $z$-component of the current at the central point~$O$ 
(see figure~\ref{FIGURE_LAYER}) is
\begin{eqnarray}
j_o\equiv(j_z)_o=(\partial_x B_y-\partial_y B_x)_o
\approx(\partial_x B_y)_o \approx B_{ext}/\delta,
\label{AMPERES_LAW}
\end{eqnarray}
where we use the estimates 
$(\partial_y B_x)_o\ll (\partial_x B_y)_o$ and 
$(\partial_x B_y)_o\approx B_{ext}/\delta$ 
at the point~$O$. The last estimate follows directly 
from the definition of $\delta$ as being the half-thickness of 
the out-of-plane current profile across the reconnection layer.

In the case of a quasi-stationary two-dimensional reconnection, 
we neglect time derivatives, and Faraday's law 
${{\bf\nabla}\times{\bf E}}=-\partial_t{\bf B}$ for
the $x$- and $y$-components of the magnetic field 
results in equations $\partial_y E_z=-\partial_t B_x=0$
and $\partial_x E_z=\partial_t B_y=0$. 
Therefore, $E_z$ is constant in space, and from 
the z-component of the generalized Ohm's law~(\ref{OHMS_LAW}) 
we obtain
\begin{eqnarray}
E_z &=& \eta j_z-V_xB_y+V_yB_x+(\omega_+^2\omega_-^2/ne)(j_xB_y-j_yB_x)
\nonumber\\
&+& \omega_+^2d_e^2\left[V_x\partial_xj_z+V_y\partial_yj_z 
+j_x\partial_xV_z+j_y\partial_yV_z \right.
\nonumber\\
&&\quad\;\left.-(\omega_+^2\omega_-^2/ne)(j_x\partial_xj_z+j_y\partial_yj_z)\right]
=\mbox{constant}.
\quad
\label{OHMS_LAW_Z}
\end{eqnarray}
The reconnection rate is determined by the value of $E_z$ at the 
central point~$O$, that is
\beq
E_z=\eta j_o.
\label{OHMS_LAW_Z_CENTER}
\eeq
We see that the electric field is balanced only by the resistive term 
$\eta j_o$ at the central point~$O$; this is because we assume isotropic 
pressure tensors in this study. To estimate $j_o$, in what follows 
we neglect time derivatives for a quasi-stationary reconnection and 
we use the symmetries of the reconnection layer.

The z-component of the momentum equation~(\ref{MOTION_LAW}) is
\beq
\rho(V_x\partial_x V_z+V_y\partial_y V_z)
+\omega_+^2d_e^2(j_x\partial_x j_z+j_y\partial_y j_z)=j_xB_y-j_yB_x. 
\nonumber
\eeq
Taking the second derivatives of this equation with respect to 
$x$ and $y$ at the point~$O$, we obtain 
\beq
\rho(\partial_xV_x)_o (\partial_{xx} V_z)_o
+\omega_+^2d_e^2(\partial_xj_x)_o(\partial_{xx} j_z)_o
=(\partial_xj_x)_o(\partial_xB_y)_o,
\nonumber
\\
\rho(\partial_yV_y)_o (\partial_{yy} V_z)_o
+\omega_+^2d_e^2(\partial_yj_y)_o(\partial_{yy} j_z)_o
=-(\partial_yj_y)_o(\partial_yB_x)_o. 
\nonumber
\eeq
Therefore,
\beq
\begin{array}{l}
(\partial_{xx}V_z)_o = -(\partial_{xy}B_z)_o
[(\partial_x B_y)_o-\omega_+^2d_e^2(\partial_{xx}j_z)_o]/\rho(\partial_yV_y)_o,\\
(\partial_{yy}V_z)_o = (\partial_{xy}B_z)_o
[(\partial_y B_x)_o+\omega_+^2d_e^2(\partial_{yy}j_z)_o]/\rho(\partial_yV_y)_o,
\end{array}
\label{V_Z_DERIVATIVES}
\eeq
where we use equations~(\ref{J_DERIVATIVES}) and the plasma 
incompressibility relation $\partial_xV_x=-\partial_yV_y$.

Next, we calculate the second derivatives of equation~(\ref{OHMS_LAW_Z}) 
with respect to $x$ and $y$ at the central point~$O$ and obtain
\beq
0 &\!=\!& \eta(\partial_{xx}j_z)_o-2[(\partial_xV_x)_o
-(\omega_+^2\omega_-^2/ne)(\partial_xj_x)_o](\partial_xB_y)_o
\nonumber
\\
&& +2\omega_+^2d_e^2[(\partial_xV_x)_o(\partial_{xx} j_z)_o
+(\partial_xj_x)_o(\partial_{xx} V_z)_o
\nonumber
\\
&& \hphantom{{}+2\omega_+^2d_e^2[{}} 
-(\omega_+^2\omega_-^2/ne)(\partial_xj_x)_o(\partial_{xx} j_z)_o],
\nonumber
\\
0 &\!=\!& \eta(\partial_{yy}j_z)_o+2[(\partial_yV_y)_o
-(\omega_+^2\omega_-^2/ne)(\partial_yj_y)_o](\partial_yB_x)_o
\nonumber
\\
&& +2\omega_+^2d_e^2[(\partial_yV_y)_o(\partial_{yy} j_z)_o
+(\partial_yj_y)_o(\partial_{yy} V_z)_o
\nonumber
\\
&& \hphantom{{}+2\omega_+^2d_e^2[{}} 
-(\omega_+^2\omega_-^2/ne)(\partial_yj_y)_o(\partial_{yy} j_z)_o].
\nonumber
\eeq
Substituting expressions~(\ref{V_Z_DERIVATIVES}) into these equations 
and using equations~(\ref{DENSITY}),~(\ref{J_DERIVATIVES}) and 
$\partial_xV_x=-\partial_yV_y$, we obtain
\begin{eqnarray}
-\eta(\partial_{xx}j_z)_o &=& 2(\partial_yV_y)_o
[(\partial_xB_y)_o-\omega_+^2d_e^2(\partial_{xx} j_z)_o]
\nonumber\\
&& \times\big[1+{\tilde\gamma}(\omega_-^2-d_e^2{\tilde\gamma}/d_i^2)\big],
\label{E_Z_PERPENDICULAR}
\\
-\eta(\partial_{yy}j_z)_o &=& 2(\partial_yV_y)_o
[(\partial_yB_x)_o+\omega_+^2d_e^2(\partial_{yy} j_z)_o]
\nonumber\\
&& \times\big[1+{\tilde\gamma}(\omega_-^2-d_e^2{\tilde\gamma}/d_i^2)\big],
\label{E_Z_PARALLEL}
\end{eqnarray}
where we introduce a useful dimensional parameter 
\begin{eqnarray}
{\tilde\gamma}\equiv \omega_+^2(\partial_{xy}B_z)_o\left/ne(\partial_y V_y)_o\right..
\label{GAMMA_TILDE}
\end{eqnarray}
In the case of electron-ion plasma ($Zm_e\ll m_i$ and 
$\omega_+^2\approx \omega_-^2\approx 1$), parameter ${\tilde\gamma}$ 
measures the relative strength of the Hall term 
$({\bf j}\times{\bf B})_z/ne$ and the ideal MHD term 
$({\bf V}\times{\bf B})_z$ inside the electron layer. 

Taking the ratio of equations~(\ref{E_Z_PERPENDICULAR}) 
and~(\ref{E_Z_PARALLEL}), we obtain
\beq
(\partial_yB_x)_o &=& (\partial_xB_y)_o(\partial_{yy}j_z)_o/(\partial_{xx}j_z)_o
-2\omega_+^2d_e^2(\partial_{yy}j_z)_o
\nonumber\\
&\approx& (B_{ext}\delta/L^2)(1+2\omega_+^2d_e^2/\delta^2),
\label{Bx_Y_INITIAL}
\eeq
where we use the estimates $(\partial_{xx}j_z)_o\approx-j_o/\delta^2$ 
and $(\partial_{yy}j_z)_o\approx-j_o/L^2$, and equation~(\ref{AMPERES_LAW}).

In equation~(\ref{OHMS_LAW_Z}), the electric field $E_z$ is balanced by the 
ideal MHD and Hall terms outside the electron layer, where the resistivity and 
electron inertia terms are insignificant. Therefore, 
\begin{eqnarray}
E_z &\approx& -V_xB_y[1-(\omega_+^2\omega_-^2/ne)j_x/V_x] 
\nonumber\\
&\approx& (\partial_y V_y)_o\delta\, B_{ext}(1+\omega_-^2{\tilde\gamma}),
\label{E_Z_X=DELTA}
\\
E_z &\approx& V_yB_x[1-(\omega_+^2\omega_-^2/ne)j_y/V_y]
\nonumber\\
&\approx& (\partial_y V_y)_o(\partial_y B_x)_oL^2(1+\omega_-^2{\tilde\gamma})
\qquad
\label{E_Z_Y=L}
\end{eqnarray}
at the points $(x\approx\delta, y=0)$ and $(x=0,y\approx L)$ 
respectively. Here we use the estimates
$j_x\approx(\partial_{xy}B_z)_o\delta$, $j_y\approx-(\partial_{xy}B_z)_oL$, 
$V_x\approx-(\partial_y V_y)_o\delta$, $V_y\approx(\partial_y V_y)_oL$,
$B_x\approx (\partial_y B_x)_oL$ and $B_y\approx B_{ext}$, and 
equation~(\ref{GAMMA_TILDE}).
The ratio of equations~(\ref{E_Z_X=DELTA}) and~(\ref{E_Z_Y=L}) gives 
\beq
(\partial_y B_x)_o\approx B_{ext}\delta/L^2\approx B_{ext}^2/j_oL^2, 
\label{Bx_Y}
\eeq
where we use equation~(\ref{AMPERES_LAW}).
Comparing this estimate with equation~(\ref{Bx_Y_INITIAL}), we find 
$\delta\gtrsim \omega_+ d_e\approx d_e$. Therefore, using 
equation~(\ref{AMPERES_LAW}), we obtain
\beq
j_o\lesssim B_{ext}/d_e
\label{J_O_LIMIT}
\eeq
and $E_z\lesssim \eta B_{ext}/d_e$~\cite{zocco_2008}.
The estimate $B_x\approx (\partial_y B_x)_oL\approx B_{ext}\delta/L$
for the value of the perpendicular magnetic field is in agreement with 
geometrical configuration of the magnetic field lines inside the
electron layer of thickness $\delta$ and length $L$. 

Combining equations~(\ref{AMPERES_LAW}),~(\ref{OHMS_LAW_Z_CENTER})
and~(\ref{E_Z_X=DELTA}), we obtain
\beq
\eta j_o^2\approx (\partial_y V_y)_o\, B_{ext}^2(1+\omega_-^2{\tilde\gamma}).
\label{ENERGY}
\eeq
This equation describes conversion of the magnetic energy into Ohmic 
heat inside the electron layer with rate~$\approx |(\partial_x u^e_x)_o|
=|(\partial_x V_x)_o-\omega_+^2(\partial_x j_x)_o/ne|
\approx (\partial_y V_y)_o(1+{\tilde\gamma})$ in the case of electron-ion 
plasma ($\omega_-^2\approx 1$)~\footnote{In the case of electron-ion 
plasma, in the upstream region outside the electron layer the magnetic 
field lines are frozen into the electron fluid and inflow with the 
electron velocity $u^e_x$.
}, 
and with rate~$\approx |(\partial_x V_x)_o|=(\partial_y V_y)_o$ in the case 
of electron-positron plasma ($\omega_-^2=0$).

Next, we use the $z$-component of Faraday's law,
$\partial_x E_y-\partial_y E_x=-\partial_t B_z=0$, where the time 
derivative is set to zero because we assume that the reconnection is 
quasi-stationary. We substitute $E_x$ and $E_y$ into this equation 
from Ohm's law~(\ref{OHMS_LAW}) and, after tedious but 
straightforward derivations, we obtain 
\beq
\eta(\partial_x j_y-\partial_y j_x)
+(\omega_+^2\omega_-^2/ne)(B_x\partial_x j_z+B_y\partial_y j_z)
\nonumber
\\
{}+V_x\partial_x B_z+V_y\partial_y B_z-B_x\partial_x V_z-B_y\partial_y V_z
\nonumber
\\
{}+\omega_+^2d_e^2[V_x(\partial_{xx} j_y-\partial_{xy} j_x)
+V_y(\partial_{xy} j_y-\partial_{yy} j_x)
\nonumber
\\
\hphantom{{}+\omega_+^2d_e^2[{}} 
+j_x(\partial_{xx} V_y-\partial_{xy} V_x)
+j_y(\partial_{xy} V_y-\partial_{yy} V_x)
\nonumber
\\
\hphantom{{}+\omega_+^2d_e^2[{}} 
-(\omega_+^2\omega_-^2/ne)j_x(\partial_{xx} j_y-\partial_{xy} j_x)
\nonumber
\\
\hphantom{{}+\omega_+^2d_e^2[{}} 
-(\omega_+^2\omega_-^2/ne)j_y(\partial_{xy} j_y-\partial_{yy} j_x)]=0.
\nonumber
\eeq
Taking the $\partial_{xy}$ derivative of this equation at the central
point~$O$ and using equations~(\ref{J_DERIVATIVES}) and~(\ref{V_Z_DERIVATIVES}), 
we obtain 
\begin{eqnarray}
0 &=&
-\eta\left[(\partial_{xyxx}B_z)_o+(\partial_{xyyy}B_z)_o\right]
+(\omega_-^2-d_e^2{\tilde\gamma}/d_i^2)
\nonumber\\
&&\qquad \times (\omega_+^2/ne)
[(\partial_x B_y)_o(\partial_{yy}j_z)_o+(\partial_y B_x)_o(\partial_{xx}j_z)_o]
\nonumber\\
&\approx& 
\eta ne(\partial_y V_y)_o{\tilde\gamma}/\omega_+^2\delta^2
\nonumber\\
&&-(\omega_-^2-d_e^2{\tilde\gamma}/d_i^2)(\omega_+^2/ne)
[j_o^2/L^2+(\partial_y B_x)_oj_o/\delta^2].
\label{E_XY_INITIAL}
\end{eqnarray}
To derive the final expression, we use equation~(\ref{GAMMA_TILDE}) 
and the estimates
$(\partial_{xyxx}B_z)_o\approx-(\partial_{xy}B_z)_o/\delta^2
\gg(\partial_{xyyy}B_z)_o$,
$(\partial_{xx}j_z)_o\approx-j_o/\delta^2$,
$(\partial_{yy}j_z)_o\approx-j_o/L^2$, 
$(\partial_x B_y)_o\approx j_o$. Using 
equations~(\ref{DENSITY}),~(\ref{S_and_V_A}),~(\ref{AMPERES_LAW})
and~(\ref{Bx_Y}), we rewrite equation~(\ref{E_XY_INITIAL}) as
\beq
\omega_-^2-d_e^2{\tilde\gamma}/d_i^2
\approx \eta L^2(\partial_y V_y)_o{\tilde\gamma}/\omega_+^2d_i^2V_A^2.
\label{E_XY}
\end{eqnarray}
Note that equations~(\ref{E_XY_INITIAL}) and~(\ref{E_XY})
result in 
\beq
0\le{\tilde\gamma}\le \omega_-^2d_i^2/d_e^2.
\label{GAMMA_TILDE_LIMITS}
\eeq

Equation~(\ref{MOTION_LAW}) for the plasma (ion) acceleration along
the reconnection layer in the $y$-direction gives 
\beq
\rho({\bf V}{\bf\nabla})V_y+\omega_+^2d_e^2({\bf j}{\bf\nabla})j_y
=-\partial_y P+j_zB_x-j_xB_z.
\label{MOTION_LAW_Y}
\eeq
Taking the $y$ derivative of this equation at the central point~$O$ 
and using equations~(\ref{DENSITY}),~(\ref{J_DERIVATIVES}) 
and~(\ref{GAMMA_TILDE}), we obtain
\begin{eqnarray}
\rho(\partial_y V_y)_o^{\,2}(1+d_e^2{\tilde\gamma}^2/d_i^2)
\approx B_{ext}^2/L^2+j_o(\partial_y B_x)_o.
\label{ACCELERATION_INITIAL}
\end{eqnarray}
In the derivation of this equation we use the estimate 
$(\partial_{yy}P)_o\approx(\partial_{yy}B_y^2/2)_{ext}\approx-B_{ext}^2/L^2$,
which reflects the fact that the pressure drop is approximately equal to 
the drop in the external magnetic field pressure. This estimate follows 
from the force balance condition for the slowly inflowing plasma across 
the layer, in analogy with the Sweet-Parker derivations~\footnote{
For a proof, integrate equation~(\ref{MOTION_LAW}) along the unclosed 
rectangular contour $(x=0,y=0)\rightarrow(x=\Delta,y=0)
\rightarrow(x=\Delta,y={\tilde y})\rightarrow(x=0,y={\tilde y})$, 
then take the limit ${\tilde y}\to 0$ and use the Taylor expansion 
in $y$ for the physical quantities that enter equation~(\ref{MOTION_LAW}). 
For details refer to~\cite{malyshkin_2005}.
}~\cite{malyshkin_2005}. 
Using equations~(\ref{S_and_V_A}) and~(\ref{Bx_Y}), and neglecting factors
of order unity, we rewrite equation~(\ref{ACCELERATION_INITIAL}) as
\begin{eqnarray}
(\partial_y V_y)_o\approx (V_A/L)(1+d_e^2{\tilde\gamma}^2/d_i^2)^{-1/2}.
\label{ACCELERATION}
\end{eqnarray}

Now we note that on the y-axis ($x=0$) equation~(\ref{MOTION_LAW_Y}) 
reduces to 
$\rho V_y \partial_y V_y=-\omega_+^2d_e^2j_y\partial_y j_y-\partial_y P+j_zB_x$.
We integrate this equation from the central point~$O$ to the 
downstream region outside of the ion layer, $x=0$ and $y\approx L_{ext}$,
where ideal MHD applies and $j_y\approx 0$. 
The plasma inertia term $\rho V_y \partial_y V_y$ integrates to 
$\rho V_y^2/2=(1/2)(B_{ext}V_y/V_A)^2$, the electron inertia term 
$\omega_+^2d_e^2j_y\partial_y j_y$ integrates to zero, the pressure 
term $-\partial_y P$ integrates to $\approx B_{ext}^2$, and the 
magnetic tension force term $j_zB_x$ integrates to 
$\approx B_{ext}^2$~\footnote{
Note that $j_z\approx j_o$ for $y\lesssim L$ and $j_z\approx 0$ for 
$y\gtrsim L$. Field $B_x\approx(\partial_y B_x)_oy\approx (B_{ext}^2/j_oL^2)y$,
see equation~(\ref{Bx_Y}). 
}. 
As a result, we find that that the eventual plasma outflow velocity is
approximately equal to the Alfven velocity, $V_y\approx V_A$, in the 
downstream region outside of the ion layer (at $y\approx L_{ext}$). 

In the end of this section, we derive an estimate for the ion layer 
half-thickness $\Delta$. In these derivations we proceed as follows. 
Outside the electron layer the electron inertia and magnetic tension 
terms can be neglected in equation~(\ref{MOTION_LAW_Y}), and we have
$\rho({\bf V}{\bf\nabla})V_y\approx-\partial_y P$. Taking
the $y$ derivative of this equation at $y=0$, we obtain 
$\rho[V_x(\partial_{xy} V_y)+(\partial_y V_y)^{\,2}]
\approx-(\partial_{yy}P)_o\approx B_{ext}^2/L^2$. Here the
term $V_x(\partial_{xy} V_y)$ is about of the same size as 
the term $(\partial_y V_y)^{\,2}$. Therefore, 
we find that $(\partial_y V_y)_{ext}\approx V_A/L$ 
outside the electron layer (but inside the ion layer). 
Next, in the upstream region outside the ion layer
ideal single-fluid MHD applies. Therefore, at 
$x\approx\Delta$ and $y=0$ equation~(\ref{OHMS_LAW_Z}) reduces to 
$E_z\approx -V_xB_y\approx-(\partial_x V_x)_{ext}\Delta\,B_{ext}
=(\partial_y V_y)_{ext}\Delta\,B_{ext}\approx V_A\Delta\,B_{ext}/L$, 
where $E_z$ is given by equation~(\ref{OHMS_LAW_Z_CENTER}).
As a result, we obtain
\begin{eqnarray}
(\partial_y V_y)_{ext}\approx V_A/L,
\qquad
\Delta\approx \eta j_oL/V_AB_{ext}.
\label{DELTA}
\end{eqnarray}

%---------------------------------------------------------------------

\section{\label{SOLUTION}
Solution for two-fluid reconnection
}

To be specific, hereafter, unless otherwise stated, 
we will focus on two-fluid reconnection in 
electron-ion plasma and will assume $Zm_e\ll m_i$, $d_e\ll d_i$ 
and $\omega_+^2=\omega_-^2=1$. In this case 
equations~(\ref{ENERGY}) and (\ref{E_XY}) reduce to 
\beq
\begin{array}{rcl}
\eta j_o^2 &\approx& (\partial_y V_y)_o\, B_{ext}^2(1+{\tilde\gamma}),
\\
1-d_e^2{\tilde\gamma}/d_i^2 &\approx& \eta L^2(\partial_y V_y)_o{\tilde\gamma}/d_i^2V_A^2,
\end{array}
\label{REDUCED_EQUATIONS}
\eeq
We solve these equations and equations~(\ref{AMPERES_LAW}), (\ref{GAMMA_TILDE}),
(\ref{Bx_Y}), (\ref{ACCELERATION}) and~(\ref{DELTA}) for unknown physical 
quantities $j_o$, $\delta$, $\Delta$, $L$, ${\tilde\gamma}$, $(\partial_y V_y)_o$, 
$(\partial_y B_x)_o$ and $(\partial_{xy}B_z)_o$. We calculate
the reconnection rate $E_z$ by using equation~(\ref{OHMS_LAW_Z_CENTER}). 
We neglect factors of order unity, and we treat the external 
field $B_{ext}$ and scale $L_{ext}$ as known parameters. 
Recall that parameter ${\tilde\gamma}$, given by equation~(\ref{GAMMA_TILDE}), 
measures the relative strength of the Hall term and the ideal MHD term 
in the z-component of Ohm's law (in the case of electron-ion plasma). 
Depending on the value of parameter ${\tilde\gamma}$, we find the 
following reconnection regimes and the corresponding solutions for
the reconnection rate.

\subsection{Slow Sweet-Parker reconnection} 

When ${\tilde\gamma}\lesssim 1$, both the Hall current and the electron 
inertia are negligible, the electrons and ions flow together, and the 
electron and ion layers have the same thickness and length. 
In this case, equations~(\ref{ACCELERATION}) and~(\ref{REDUCED_EQUATIONS}) 
become $(\partial_y V_y)_o \approx V_A/L$, 
$\eta j_o^2 \approx (\partial_y V_y)_o\, B_{ext}^2$ and 
$1 \approx \eta L^2(\partial_y V_y)_o{\tilde\gamma}/d_i^2V_A^2$ respectively.
As a result, we obtain the Sweet-Parker solution~\cite{sweet_1958,parker_1963}, 
\begin{eqnarray}
\begin{array}{l}
1\ll S=V_AL_{ext}/\eta \lesssim L_{ext}^2/d_i^2,
\\
{\tilde\gamma} \approx V_Ad_i^2/\eta L_{ext} = Sd_i^2/L_{ext}^2,
\\
E_z \approx \eta^{1/2}V_A^{1/2}B_{ext}/L_{ext}^{1/2} = V_AB_{ext}/S^{1/2},
\\
j_o \approx V_A^{1/2}B_{ext}/\eta^{1/2}L_{ext}^{1/2} = S^{1/2}B_{ext}/L_{ext},
\\
\delta\approx\Delta \approx \eta^{1/2}L_{ext}^{1/2}/V_A^{1/2} = L_{ext}/S^{1/2}\gtrsim d_i, 
\\
L \approx L_{ext},
\\
(\partial_y V_y)_o \approx (\partial_y V_y)_{ext}\approx V_A/L_{ext},
\\
(\partial_y B_x)_o \approx \eta^{1/2}B_{ext}/V_A^{1/2}L_{ext}^{3/2} = B_{ext}/L_{ext}S^{1/2},
\\
(\partial_{xy} B_z)_o \approx V_AB_{ext}d_i/\eta L_{ext}^2 = SB_{ext}d_i/L_{ext}^3,
\end{array}
\label{SLOW}
\end{eqnarray}
where the Lundquist number $S\gg 1$ is defined by equation~(\ref{S_and_V_A}).
The condition $S\lesssim L_{ext}^2/d_i^2$ is obtained from 
${\tilde\gamma}\lesssim 1$. From this condition for $S$ we find that Sweet-Parker 
reconnection takes place when $d_i$ is less than the Sweet-Parker layer thickness,  
$d_i\lesssim L_{ext}/S^{1/2}$, which is a result observed in numerical
simulations~\cite{drake_2006,yamada_2009,zweibel_2009}. 
Note that the quadrupole field is small in the Sweet-Parker reconnection case,
$B_z\approx (\partial_{xy} B_z)_oL\delta\approx (S^{1/2}d_i/L_{ext})B_{ext}\lesssim B_{ext}$,
and the ion and electron outflow velocities are approximately equal to 
the Alfven velocity, $V_y\approx (\partial_y V_y)_oL\approx V_A$
\cite{yamada_2009,zweibel_2009}.

Now, let us for a moment consider the case of reconnection in 
electron-positron plasma. In this case $d_e=d_i$, $\omega_+^2=1/2$, 
$\omega_-^2=0$ and equation~(\ref{GAMMA_TILDE_LIMITS}) gives 
${\tilde\gamma}=0$. This result represents an absence of the quadrupole field 
$B_z$ [refer to equation~(\ref{GAMMA_TILDE})], which is known from numerical 
simulations~\cite{bessho_bhattacharjee_2005,daughton_karimabadi_2007,drake_etal_2008}. 
Therefore, our model predicts the slow Sweet-Parker reconnection 
solution for reconnection in electron-positron plasmas, which is 
in disagreement with the results of kinetic numerical 
simulations~\cite{bessho_bhattacharjee_2005,daughton_karimabadi_2007,drake_etal_2008}.
A likely reason for this discrepancy is that our model
neglects pressure tensor anisotropy, which plays an important 
role in reconnection in electron-positron plasma.

\subsection{Transitional Hall reconnection}
 
When $1\lesssim{\tilde\gamma}\lesssim d_i/d_e$, 
the Hall current is important but the electron inertia is negligible. 
In this case, equations~(\ref{ACCELERATION}) and~(\ref{REDUCED_EQUATIONS}) 
become $(\partial_y V_y)_o \approx V_A/L$, 
$\eta j_o^2 \approx (\partial_y V_y)_o\, B_{ext}^2{\tilde\gamma}$ and
$1 \approx \eta L^2(\partial_y V_y)_o{\tilde\gamma}/d_i^2V_A^2$. 
As a result, we obtain the following solution:
$1 \lesssim {\tilde\gamma} \approx d_i^2V_A/\eta L = Sd_i^2/LL_{ext} \lesssim d_i/d_e$,
$E_z \approx (d_i/L)V_AB_{ext}$,
$j_o \approx d_iV_AB_{ext}/\eta L = Sd_iB_{ext}/LL_{ext}$,
$\delta \approx \eta L/d_iV_A = LL_{ext}/Sd_i$,
$\Delta\approx d_i$,
$(\partial_y V_y)_o \approx (\partial_y V_y)_{ext}\approx V_A/L$,
$(\partial_y B_x)_o \approx \eta B_{ext}/d_iV_AL = B_{ext}L_{ext}/Sd_iL$,
$(\partial_{xy} B_z)_o \approx d_iV_AB_{ext}/\eta L^2 = Sd_iB_{ext}/L^2L_{ext}$.
These results are in agreement with earlier theoretical 
findings~\cite{malyshkin_2008,cowley_1985,bhattacharjee_2001,simakov_2008}.

Condition $1\lesssim {\tilde\gamma}\lesssim d_i/d_e$
gives $Sd_ed_i/L_{ext}\lesssim L\lesssim Sd_i^2/L_{ext}$
for the electron layer length $L$. Unfortunately, in our model, 
the exact value of $L$ cannot be estimated in the Hall reconnection regime. 
In theoretical studies~\cite{malyshkin_2008,cowley_1985,simakov_2008} 
length $L$ was essentially treated as a fixed parameter. 
Here, we take a different approach and make a conjecture that 
the Hall reconnection regime describes a transition from the slow 
Sweet-Parker reconnection to the fast collisionless reconnection 
(presented in the next section). 
Numerical simulations and laboratory experiments have demonstrated that 
this transition happens when the ion inertial length is approximately 
equal to the Sweet-Parker layer thickness, 
$d_i\approx L_{ext}/\sqrt{S}$~\cite{drake_2006,yamada_2009,zweibel_2009,
huba_2004,murphy_2008}. 
Therefore, our conjecture leads to the following solution for 
the Hall reconnection regime:
\begin{eqnarray}
\begin{array}{l}
S=V_AL_{ext}/\eta\approx L_{ext}^2/d_i^2,
\\
L_{ext} \gtrsim L \gtrsim d_eL_{ext}/d_i,
\\
{\tilde\gamma} \approx L_{ext}/L,
\\
E_z \approx (d_i/L)V_AB_{ext},
\\
j_o \approx B_{ext}L_{ext}/d_iL,
\\
\delta \approx d_iL/L_{ext}\gtrsim d_e,
\\
\Delta\approx d_i \gtrsim \delta,
\\
(\partial_y V_y)_o \approx (\partial_y V_y)_{ext}\approx V_A/L,
\\
(\partial_y B_x)_o \approx B_{ext}d_i/LL_{ext},
\\
(\partial_{xy} B_z)_o \approx B_{ext}L_{ext}/d_iL^2.
\end{array}
\label{HALL}
\end{eqnarray}
It is noteworthy that, in the Hall reconnection regime, the typical value 
of the quadrupole field is comparable to the reconnecting field value, 
$B_z\approx (\partial_{xy} B_z)_oL\delta\approx B_{ext}$.
The typical value of the ion outflow velocity is equal to 
the Alfven velocity, 
$V_y\approx (\partial_y V_y)_oL\approx V_A$. 
To estimate the typical value of the electron outflow velocity, we use 
equations~(\ref{E_I_VELOCITIES}), (\ref{DENSITY}), 
(\ref{J_DERIVATIVES}) and (\ref{HALL}), and find
$u^e_y\approx V_y-(m_i/Ze\rho)j_y=V_y-(d_iV_A/B_{ext})j_y
\approx V_A+(d_iV_A/B_{ext})(\partial_{xy} B_z)_oL
\approx V_A(L_{ext}/L)\gtrsim V_A$.

As the electron layer length $L$ decreases from its 
maximal value $L\approx L_{ext}$ to its minimal value 
$L\approx d_eL_{ext}/d_i$, the transitional Hall reconnection 
solution~(\ref{HALL}) changes from the slow Sweet-Parker 
solution~(\ref{SLOW}) to the fast collisionless reconnection 
solution presented below 
[see equations~(\ref{S_FAST})-(\ref{Bz_XY_FAST}) and 
table~\ref{TABLE_SOLUTION}].

\subsection{Fast collisionless reconnection}
 
When $d_i/d_e\lesssim{\tilde\gamma}<d_i^2/d_e^2$ [compare to
equation~(\ref{GAMMA_TILDE_LIMITS})], the electron 
inertia and the Hall current are important inside the electron 
layer and the ion layer respectively. In this case, 
equations~(\ref{ACCELERATION}) and~(\ref{REDUCED_EQUATIONS}) become 
$(\partial_y V_y)_o{\tilde\gamma} \approx d_iV_A/d_eL$,
$\eta j_o^2 \approx (\partial_y V_y)_o\, B_{ext}^2{\tilde\gamma}$ and
$1-d_e^2{\tilde\gamma}/d_i^2 \approx \eta L^2(\partial_y V_y)_o{\tilde\gamma}/d_i^2V_A^2$.
As a result, taking into consideration equation~(\ref{J_O_LIMIT}), 
we obtain the following solution:
\begin{eqnarray}
L_{ext}/d_e \ll S=V_AL_{ext}/\eta \lesssim L_{ext}^2/d_ed_i,
\label{S_FAST}
\\
d_i/d_e \lesssim {\tilde\gamma} < d_i^2/d_e^2,
\label{GAMMA_FAST}
\\
E_z \approx \eta B_{ext}/d_e=(L_{ext}/Sd_e)V_AB_{ext}
\nonumber\\
\hphantom{E_z} \approx (\Delta/L)V_AB_{ext}\approx (d_i/L)V_AB_{ext},
\label{Ez_FAST}
\\
j_o \approx B_{ext}/d_e,
\label{J_O_FAST}
\\
\delta\approx d_e,
\label{DELTA_FAST}
\\
\Delta \approx d_i \gg \delta,
\label{DELTA_ION_FAST}
\\
L \approx V_Ad_ed_i/\eta = Sd_ed_i/L_{ext},
\label{L_FAST}
\\
(\partial_y V_y)_o \approx \eta/d_e^2{\tilde\gamma}
=V_AL_{ext}/Sd_e^2{\tilde\gamma} \lesssim V_A/L,
\label{Vy_Y_FAST}
\\
(\partial_y V_y)_{ext} \approx \eta/d_ed_i=V_AL_{ext}/Sd_ed_i\approx V_A/L,
\label{Vy_Y_EXT_FAST}
\\
(\partial_y B_x)_o \approx B_{ext}\eta^2/V_A^2d_ed_i^2
=B_{ext}L_{ext}^2/S^2d_ed_i^2,\quad
\label{Bx_Y_FAST}
\\
(\partial_{xy} B_z)_o \approx B_{ext}\eta/V_Ad_e^2d_i
=B_{ext}L_{ext}/Sd_e^2d_i.
\label{Bz_XY_FAST}
\end{eqnarray}
Here the limits on the Lundquist number given in equation~(\ref{S_FAST}),
$L_{ext}/d_e \ll S \lesssim L_{ext}^2/d_ed_i$, are obtained from the 
conditions $E_z\ll V_AB_{ext}$ (slow quasi-stationary reconnection) and 
$L\lesssim L_{ext}$ (the electron layer length cannot exceed 
the ion layer length). 
Except for the definition of the reconnecting field $B_{ext}$, 
equations~(\ref{Ez_FAST})-(\ref{DELTA_FAST}) and~(\ref{L_FAST}) essentially 
coincide with the results obtained in~\cite{zocco_2008} for a 
model of electron MHD (EMHD) reconnection.
The collisionless reconnection rate, given by equation~(\ref{Ez_FAST}), 
is much faster than the Sweet-Parker rate $E_z\approx V_AB_{ext}/\sqrt{S}$ 
[see equations~(\ref{SLOW})]. 

Note that the value of ${\tilde\gamma}$ or, alternatively, the 
value of the ion acceleration rate 
$(\partial_y V_y)_o\approx \eta/d_e^2{\tilde\gamma}$ at the 
point~$O$ cannot be determined exactly. This is because in the 
plasma momentum equation~(\ref{MOTION_LAW_Y}), the magnetic tension 
and pressure forces are balanced by the electron inertia term 
$d_e^2({\bf j}{\bf\nabla})j_y$ inside the electron layer.
The ion inertia term $\rho({\bf V}{\bf\nabla})V_y$ can be of the 
same order or smaller, resulting in the upper limit 
$(\partial_y V_y)_o\lesssim V_A/L$. In other words, inside 
the electron layer the magnetic energy is converted into the 
kinetic energy of the electrons (and into Ohmic heat), while the ion 
kinetic energy can be considerably smaller. Therefore, the ion 
outflow velocity can be significantly less than $V_A$ in the 
downstream region outside the electron layer (at $y\approx L$). 
At the same time, the electron outflow velocity is much larger than 
$V_A$ and is approximately equal to the electron Alfven velocity,
$u^e_y\approx (m_i/Ze\rho)j_y=(d_iV_A/B_{ext})(\partial_{xy} B_z)_oL
\approx d_iV_A/d_e\approx V_{eA}\equiv B_{ext}/\sqrt{nm_e}\gg V_A$.
However, further in the downstream region, at $y\gtrsim L$, as the 
electrons gradually decelerate, their kinetic energy is converted
into the ion kinetic energy. As a result, the eventual ion 
outflow velocity becomes $\approx V_A$, as was estimated in the 
end of Section~\ref{RECONNECTION_EQUATIONS}. 
These results emphasize the critical role that electron inertia 
plays in the plasma momentum equation~(\ref{MOTION_LAW}). These results 
also agree with simulations~\cite{daughton_2007}, which found the ion 
outflow velocity to be significantly less than $V_A$ in the downstream 
region outside of the electron layer, and found acceleration of ions 
further downstream (in the decelerating electron outflow jets).

Our theoretical results for collisionless reconnection are in good 
agreement with numerical simulations and/or laboratory 
experiments~\footnote{
Even though reconnection rate~(\ref{Ez_FAST}) is proportional to resistivity,
we still use the standard term ``collisionless reconnection'' because in the 
fast reconnection regime $\eta$ should be viewed as the effective 
resistivity, which is to be calculated from the kinetic theory.
}.
Indeed, the estimates 
$\Delta\approx d_i$ for the ion layer thickness, 
$\delta\approx d_e$ for the electron layer thickness, 
$B_z\approx (\partial_{xy} B_z)_o\delta L\approx B_{ext}$ for
the quadrupole field, and 
$u_y^e\approx V_{eA}\equiv B_{ext}/\sqrt{nm_e}$ for the electron 
outflow velocity agree with 
simulations~\cite{drake_2006,yamada_2009,zweibel_2009,daughton_2006,
fujimoto_2006,daughton_2007,shay_2007}.
The estimates $\Delta\approx d_i$ and $B_z\approx B_{ext}$ also 
agree with experiment~\cite{yamada_2009}. However, 
the experimentally measured thickness of the electron layer is about 
eight times larger than our theoretical model and numerical simulations 
predict~\cite{ren_yamada_2008,ji_daughton_2008}. This discrepancy
can be due to three-dimensional geometry effects and plasma instabilities
that may play an important role in the 
experiment~\cite{yamada_2009,ji_daughton_2008}. 

Our results are also in a qualitative agreement with recent 
numerical findings of an inner electron  dissipation layer and 
of electron outflow jets that extend into the ion 
layer~\cite{daughton_2006,fujimoto_2006,daughton_2007,shay_2007}. 
We note that the estimated electron layer length $L\approx V_Ad_ed_i/\eta$ 
is generally much larger than both the electron layer thickness 
$\delta\approx d_e$ and the ion layer thickness $\Delta\approx d_i$, 
which is consistent with numerical 
simulations~\cite{daughton_2006,fujimoto_2006,daughton_2007}.
However, if resistivity $\eta$ becomes anomalous and considerably enhanced 
over the Spitzer value, then $L$ can theoretically become of order of $d_i$ 
and the reconnection rate can become comparable to the Alfven rate 
$V_AB_{ext}$, which is also observed in numerical 
simulations~\cite{huba_2004,shay_2007}.

Unfortunately, a detailed quantitative comparison of our theoretical results 
to the results of kinetic numerical simulations is not possible because these 
simulations do not explicitly specify constant resistivity $\eta$. 
In addition, in the simulations the anisotropy of the electron pressure 
tensor anisotropy was found to play an important role inside the electron 
layer and in the electron outflow 
jets~\cite{daughton_2007,shay_2007}. In contrast, in the present study we
assume an isotropic pressure, and the electrons are coupled to 
the field lines everywhere outside the electron layer (including the jets).

In our model, the electric field $E_z$ is supported by the Hall term 
$({\bf j}\times{\bf B})_z/ne$ in the downstream region $L\lesssim y\lesssim L_{ext}$.
Therefore, in the collisionless reconnection regime, our model predicts 
an existence of Hall-MHD Petschek shocks that are attached to the two ends 
of the electron layer and separate the two electron outflow jets and the 
surrounding plasma. Note that, for electron-ion plasma ($Zm_e\ll m_i$), the 
ideal MHD and Hall terms in Ohm's law~(\ref{OHMS_LAW_INITIAL}) can be 
combined together as 
$-{\bf V}\times{\bf B}+(m_i/Ze\rho){\bf j}\times{\bf B}=-{\bf u^e}\times{\bf B}$,
where ${\bf u^e}$ is the electron velocity given by equation~(\ref{E_I_VELOCITIES}).
Therefore, all results for the Hall-MHD Petschek shocks can be obtained
from the corresponding results derived for the standard MHD Petschek shocks 
by replacing the plasma velocity ${\bf V}$ with the electron velocity ${\bf u^e}$. 
In particular, the parallel components of the magnetic field and 
electron velocity jump across the Hall-MHD Petschek shocks, the velocity
of the shocks is $\approx |u^e_x|\approx (m_i/Ze\rho)|j_x|
\approx (d_iV_A/B_{ext})(\partial_{xy} B_z)_o\delta\approx V_AL_{ext}/Sd_e\ll V_A$, 
and the opening angle between the shocks is 
$\approx B_x/B_y\approx (\partial_y B_x)_oL/B_{ext}\approx L_{ext}/Sd_i\ll 1$. 
Shocks were indeed observed in numerical simulations~\cite{arber_2006}. 
However, in these simulations a spatially localized anomalous resistivity 
was prescribed, resulting in a short layer length, while in our study 
resistivity $\eta$ is assumed to be constant.

\begin{table}
\caption{Solution for two-fluid reconnection\label{TABLE_SOLUTION}}
\smallskip
\begin{tabular}{|l|c|c|c|}
\hline
 &
slow Sweet-Parker &
Hall &
fast 
\\
\hline
%$S\!=\!V_AL_{ext}/\eta$ & 
$S$ & 
$1\ll S \lesssim L_{ext}^2/d_i^2$ &
$L_{ext}^2/d_i^2$ &
$L_{ext}/d_e \ll S \lesssim L_{ext}^2/d_ed_i$
\\
\hline
${\tilde\gamma}$ & 
$Sd_i^2/L_{ext}^2$ &
$L_{ext}/L$ &
$d_i/d_e \lesssim {\tilde\gamma} < d_i^2/d_e^2$
\\
\hline
$E_z$ & 
$V_AB_{ext}/S^{1/2}$ &
$(d_i/L)V_AB_{ext}$ &
$(L_{ext}/Sd_e)V_AB_{ext}$
\\
 &
 &
 &
$\approx (d_i/L)V_AB_{ext}$
\\
\hline
$j_o$ &
$S^{1/2}B_{ext}/L_{ext}$ &
$B_{ext}L_{ext}/d_iL$ &
$B_{ext}/d_e$
\\
\hline
$\delta$ &
$L_{ext}/S^{1/2}\gtrsim d_i$ &
$d_iL/L_{ext}\gtrsim d_e$ &
$d_e$
\\
\hline
$\Delta$ &
$L_{ext}/S^{1/2}\approx \delta$ &
$d_i \gtrsim \delta$ &
$d_i \gg \delta$
\\
\hline
$L$ &
$L_{ext}$ &
$L_{ext} \gtrsim L \gtrsim d_eL_{ext}/d_i$ &
$Sd_ed_i/L_{ext}$
\\
\hline
$(\partial_y V_y)_o$ &
$V_A/L$ &
$V_A/L$ &
$V_AL_{ext}/Sd_e^2{\tilde\gamma} \lesssim V_A/L$
\\
\hline
$(\partial_y V_y)_{ext}$ &
$V_A/L$ &
$V_A/L$ &
$V_AL_{ext}/Sd_ed_i\approx V_A/L$
\\
\hline
$(\partial_y B_x)_o$ & 
$B_{ext}/L_{ext}S^{1/2}$ &
$B_{ext}d_i/LL_{ext}$ &
$B_{ext}L_{ext}^2/S^2d_ed_i^2$
\\
\hline
$(\partial_{xy} B_z)_o$ &
$SB_{ext}d_i/L_{ext}^3$ &
$B_{ext}L_{ext}/d_iL^2$ &
$B_{ext}L_{ext}/Sd_e^2d_i$
\\
\hline
\end{tabular}
\end{table}

%---------------------------------------------------------------------

\section{\label{DISCUSSION}
Discussion
}

\begin{figure}[t]
\vspace{5.0truecm}
\includegraphics{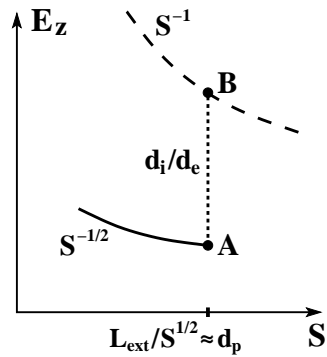}
\caption{Schematic plot of the reconnection rate $E_z$ versus 
the Lundquist number $S$ in the slow Sweet-Parker (solid line), 
transitional Hall (dotted line), and fast collisionless (dashed line) 
reconnection regimes.
}
\label{SOLUTION_PICTURE}
\end{figure}

The solution for two-fluid reconnection is summarized in 
table~\ref{TABLE_SOLUTION}. This table includes solution formulas for 
three reconnection regimes: the slow Sweet-Parker reconnection regime, 
the transitional Hall reconnection regime, and the fast collisionless 
reconnection regime. 
The reconnection rates for these three regimes are respectively shown 
by the solid, dotted and dashed lines in figure~\ref{SOLUTION_PICTURE}. 

It is well known that resistivity $\eta$ can be considerably enhanced 
by current-driven plasma instabilities~\cite{yamada_2009,zweibel_2009,kulsrud_2001}. 
Because the collisionless reconnection rate $E_z\approx\eta B_{ext}/d_e$ 
is proportional to the resistivity [see equation~(\ref{Ez_FAST})], 
this rate can increase significantly as well.
As a result, we propose the following possible theoretical explanation 
for the two-stage reconnection behavior (fast and slow) that is 
frequently observed in cosmic and laboratory plasma systems  
undergoing reconnection processes.

During the first stage, such a system is in the very slow Sweet-Parker 
reconnection regime, during which magnetic energy is slowly built up and 
stored in the system. The magnetic energy and electric currents build 
up, the field strength increases and the resistivity decreases~\cite{uzdensky_2007}. 
As a result, the Lundquist number $S$ increases and the system moves to the right 
along the solid line in figure~\ref{SOLUTION_PICTURE}. 

When the Lundquist number $S$ becomes comparable to $L_{ext}^2/d_i^2$ and 
the thickness of the current layer $L_{ext}/S^{1/2}$ becomes comparable to 
$d_i$, the system reaches point~A in figure~\ref{SOLUTION_PICTURE}. Next 
the system goes into the transitional Hall reconnection regime and quickly 
moves up along the vertical dotted line in figure~\ref{SOLUTION_PICTURE}. 
During this transition, the length of the electron layer shrinks from 
$\approx L_{ext}$ to $\approx(d_e/d_i)L_{ext}$, the electron layer 
thickness decreases from $\approx d_i$ to $\approx d_e$, and both the 
electric current and the reconnection rate increase by a factor 
$\approx d_i/d_e\gg 1$. The system ends up in the fast collisionless 
reconnection regime at point~B in figure~\ref{SOLUTION_PICTURE}. 

Because of the considerable increase in the electric current 
during the Hall reconnection transition from point~A to point~B, plasma 
instabilities develop, and, consequently, resistivity $\eta$ becomes 
anomalous and rises in value. 
As a result, the reconnection rate $E_z\approx\eta B_{ext}/d_e$ increases, 
the Lundquist number $S=V_AL_{ext}/\eta$ and electron layer length 
$L\approx V_Ad_ed_i/\eta$ decrease, and the system moves from point~B 
to the left along the dashed line in figure~\ref{SOLUTION_PICTURE}. 
The system enters the second stage characterized by a rapid release 
of the accumulated magnetic energy. Even though our theoretical model 
is stationary, assumes constant resistivity and cannot describe this 
stage in detail, the physical mechanism of slow and fast 
reconnection outlined above is self-consistent and may take place 
in nature.

%---------------------------------------------------------------------

\section*{Acknowledgments}

I would like to thank F.~Cattaneo, A.~Das, H.~Ji, D.~Lecoanet, R.~Kulsrud, 
J.~Mason, A.~Obabko, D.~Uzdensky and M.~Yamada for useful discussions. 
This study was supported by the NSF Center for Magnetic 
Self-Organization (CMSO), NSF award \#PHY-0821899.

%---------------------------------------------------------------------

%---------------------------------------------------------------------

\end{document}